\documentclass[aps,prd,nofootinbib,twocolumn,floatfix,superscriptaddress]{revtex4}
\usepackage{graphicx}
\usepackage{dcolumn}
\usepackage{times,mathptm}
\newcommand{\beq}{\begin{equation}}
\newcommand{\eeq}{\end{equation}}
\newcommand{\bea}{\begin{eqnarray}}
\newcommand{\eea}{\end{eqnarray}}

\renewcommand{\d}{\delta}
\renewcommand{\l}{\lambda}

\renewcommand{\b}{\beta}

\newcommand{\g}{\gamma}

\newcommand{\m}{\mu}

\newcommand{\s}{\sigma}

\newcommand{\D}{{\cal D}}

\newcommand{\oh}{{\textstyle{\frac{1}{2}}}}

\newcommand{\oq}{{\textstyle{\frac{1}{4}}}}

\newcommand{\dg}{\dagger}
\newcommand{\non}{\nonumber}

\newcommand{\rf}[1]{(\ref{#1})}
\newcommand{\ra}{\rightarrow}
\newcommand{\pa}{\partial}

\begin{document}

\title{Confinement from Gauge Invariance in 2+1 Dimensions} 

\author{J. Greensite}

\affiliation{The Niels Bohr Institute, Blegdamsvej 17, 
DK-2100 Copenhagen \O, Denmark}
\affiliation{Physics and Astronomy Dept., San Francisco State
University, San Francisco, CA~94132, USA}

\author{{\v S}. Olejn\'{\i}k}
\affiliation{Institute of Physics, Slovak Academy
of Sciences, SK--845 11 Bratislava, Slovakia}
 
\date{\today}
\begin{abstract}
   It is shown, in $D=2+1$ dimensions, that by merely imposing non-abelian 
gauge invariance on the temporal gauge ground state wavefunctional of an abelian gauge theory, a 
confining state is obtained.
\end{abstract}

\maketitle
%
%

     Confinement is a property of the vacuum of quantized 
non-abelian gauge theories. In the Hamiltonian formulation of quantized
Yang-Mills theory, in $D=d+1$ dimensions and temporal gauge, the vacuum 
is the ground state solution $\Psi_{0}[A]$ of the time-independent 
Schr\"odinger equation
\beq
             H\Psi_{0} = E_{0} \Psi_{0}
\eeq
where
\beq
           H = \int d^{d}x \left\{ - \oh{\d^{2} \over \d A_{k}^{a}(x)^{2}} + \oq F_{ij}^{a}(x)^{2} \right\}
\eeq
Physical states in temporal gauge, in SU(2) gauge theory, are required to satisfy the Gauss Law constraint
\beq
 \Bigl(\d^{ac} \pa_k + g \epsilon^{abc} A_k^b\Bigr){\d \over \d A_k^c}
           \Psi = 0
\eeq
which is the condition for invariance under infinitesimal gauge transformations.
In strong-coupling lattice gauge theory, the ground state can be solved for systematically,
order-by-order in powers of $1/g^{4}$ \cite{Me2}
\beq
            \Psi_{0}[U] = \exp\left[ -\sum_{n=1}^{\infty} \left({1\over g^{4}}\right)^{n} R_{n}[U] \right]
\eeq
where  the $R_{n}[U]$ are gauge-invariant expressions formed from
Wilson loops on the lattice.  
If the sum in the exponent is truncated at order $n=M$, the resulting state solves the time-independent
Schr\"odinger equation to $M$-th order in $1/g^{4}$, but satisfies the physical state condition exactly.

    Similarly, we may guess that at weak couplings there is also a weak-coupling expansion for the ground state.
This would have the form (in continuum notation, with lattice regularization implicit)
\beq
          \Psi_{0}[A] = \exp\left[ -\sum_{n=0}^{\infty} g^{2n} Q_{n}[A] \right]
\eeq
where the $Q_{n}[A]$ are gauge invariant, and the approximate solution, with the sum truncated to
some order in $g^{2}$, satisfies the Schr\"odinger equation to that order.   It is awkward to build the
$Q_{n}[A]$ out of Wilson loops at weak couplings; the strong-coupling series suggests that such a construction
would involve summation over infinite sets of loops in the continuum limit.  Instead, we conjecture that
that the building blocks of the $Q_{n}[A]$ are the covariant derivatives, and their associated covariant Green's 
functions. If that is so, then the choice of $Q_0$ is essentially unique. We know that in the $g=0$ limit, 
the ground state wavefunctional is just
\bea
          \Psi_{0}[A] &=& \exp\left[-\oq\int d^{d}x d^{d}y ~\Bigl(\pa_{i}A^{a}_{j}(x) - \pa_{j}A^{a}_{i}(x)\Bigr) \right.
\non \\
     & & \left. \qquad \times \left({\d^{ab}\over \sqrt{-\nabla^{2}}}\right)_{xy}  
            \Bigl(\pa_{i}A^{b}_{j}(y) - \pa_{j}A^{b}_{i}(y)\Bigr)  \right]
\label{free}
\eea
Then a state which solves the Schr\"odinger equation to zeroth order in the coupling $g$,
but satisfies the Gauss Law constraint exactly, is obtained by simply replacing ordinary derivatives
by covariant derivatives; i.e.\  
\beq
          \Psi_{0}[A] = \exp\left[-\oq\int d^{d}x d^{d}y ~F^{a}_{ij}(x) 
          \left({1\over \sqrt{-\D^{2}}}\right)^{ab}_{xy} F^{b}_{ij}(y) \right]
\label{state}
\eeq
where $\D_{k}$ is the covariant derivative in the adjoint representation of the
gauge group.  Field strengths are of course commutators of covariant derivatives, so this expression
can be regarded as a functional of covariant derivatives only.
  
    The non-confining ground-state solution of the free field theory, eq.\ \rf{free}, is well-known.
Many years ago, it was suggested by one of us \cite{Me1}
that, at large distance scales, 
the effective Yang-Mills vacuum state is simply~\footnote{See also Halpern \cite{Marty} and Mansfield \cite{Mansfield}.} 
\beq
         \Psi_{0}^{eff}[A] \approx \exp\left[- \m \int d^{d}x ~F^{a}_{ij}(x) F^{a}_{ij}(x) \right] 
\label{eff}
\eeq   
This vacuum state has the property of \emph{dimensional reduction}; i.e.\ the computation of
a space-like loop in $d+1$ dimensions looks just like the computation of a loop in a $d$-dimensional
Euclidean Yang-Mills theory.  If both the vacua  $\Psi_{0}^{(3)}$ of the $3+1$ dimensional theory, and $\Psi_{0}^{(2)}$
of the $2+1$ dimensional theory have this form,
then the computation of a large planar
Wilson loop in 3+1 dimensions reduces, in two steps, to a computation in two-dimensional Yang-Mills theory, i.e. 
\bea
           W(C) &=& \langle \mbox{Tr}[U(C)]\rangle^{D=4} 
                      = \langle \Psi^{(3)}_{0}|\mbox{Tr}[U(C)]|\Psi^{(3)}_{0}\rangle
\non \\
                        &\sim&  \langle \mbox{Tr}[U(C)]\rangle^{D=3} 
                      = \langle \Psi^{(2)}_{0}|\mbox{Tr}[U(C)]|\Psi^{(2)}_{0}\rangle
\non \\
                        &\sim& \langle \mbox{Tr}[U(C)]\rangle^{D=2} 
\label{dimred}
\eea
In $D=2$ Euclidean dimensions the Wilson loop can be calculated analytically, and we know that there is an
area-law falloff.   

      In Hamiltonian lattice gauge theory it is possible to calculate the vacuum state systematically
in powers of $1/g^{4}$ \cite{Me2}, and the result agrees with the above conjecture for the continuum theory.
Further support for this form of the ground state wavefunctional was obtained via lattice Monte Carlo
simulations \cite{Me3}.  In an interesting paper from 1998, Karabali, Kim, and Nair \cite{Nair} came up 
with a strong-coupling expansion for the vacuum state of the continuum Yang-Mills 
theory in $D=2+1$ dimensions (see also Leigh et al.\ \cite{Rob}).  This state, to leading order in $1/g^{4}$, 
also agrees with eq.\ \rf{eff}.    
 
      If we take eq.\ \rf{state} as the zeroth-order approximation to the weak-coupling ground state,  and if
the kernel $1/\sqrt{-\D^{2}}$ would have a finite range, then the effective ground state at scales much larger than
this range would have the dimensional reduction form \rf{eff}.  In order to make this statement a little more
precise in $D=2+1$ dimensions, consider the decomposition of the field strength at fixed time, in the $x-y$ plane, 
into gauge-covariant and gauge-invariant factors:
\bea
            F_{12}^{a}(x) &=& \phi^{a}(x) f(x)  
\non \\
          \phi^{a}(x) &=& {F_{12}^{a}(x)\over \sqrt{F_{12}^{c}F_{12}^{c}}}
                       ~~,~~ f(x) = \sqrt{F_{12}^{c}(x)F_{12}^{c}(x)}
\eea
In SU(2) gauge theory $\phi^{a}(x)$ is a unit 3-vector in color space, which can be made to point in any direction,
at any position $x$, by a suitable choice of gauge.  In the $x-y$ plane, a gauge-invariant distinction between
high and low-frequency field-strength components can be made in terms of the Fourier transform $f(k)$ of
$f(x)$.   For some reference momentum scale $\s$, define  
\bea
                f^{low}(x) &=& \int_{|k|<\s} {d^{2}k \over (2\pi)^{2}}~ f(k) e^{ikx}
\non \\
                f^{high}(x) &=& \int_{|k|\ge \s} {d^{2}k \over (2\pi)^{2}}~ f(k) e^{ikx}
\non \\
                F^{low/high,a}_{12}(x) &=&  \phi^{a}(x) f^{low/high}(x)  
\eea
We will also write $\Psi_{0} = e^{-R}$ with
\beq
             R = \oh\int  d^{2}x d^{2}y ~ f(x) \phi^{a}(x) \left({1\over \sqrt{-\D^{2}}}\right)^{ab}_{xy} \phi^{b}(y) f(y)
\eeq
Now, in the spirit of self-consistent field approaches, let us approximate the gauge-invariant expression
$\phi (1/\sqrt{-\D^{2}}) \phi$ by its expectation value
\beq
           S(x-y) = \left\langle \phi^{a}(x) \left({1\over \sqrt{-\D^{2}}}\right)^{ab}_{xy} \phi^{b}(y) \right\rangle
\eeq
If the covariant vacuum kernel $1/\sqrt{-\D^{2}}$ has a finite range $l_{K}$, then $S(x-y)$ has the same range,
and its low-momentum components are approximately constant, i.e.\ $S(k) \approx \g$, for $|k| \ll \pi/l_{K}$.  If we
also choose $\s \ll \pi/l_{K}$, we find
\newpage
\bea
          R &=& {\g\over 2} \int d^{2}x ~ F_{12}^{low,a}(x) F_{12}^{low,a}(x) 
\non \\
      & & +  \oh\int d^{2}x d^{2}y ~ f^{high}(x) S(x-y) f^{high}(y)
\eea
Then $f^{high}(x)$ decouples from $f^{low}(x)$, 
and the dependence of the wavefunctional on the long range, ``low-momentum" component $F^{low,a}_{12}(x)$ 
of the field strength is given by the dimensional reduction form \rf{eff}.   
The fluctuations of $F^{low,a}_{12}(x)$  are then essentially local, since in $d=2$ dimensions there is no
Bianchi identity among field strengths which would induce a correlation.    On the other hand, with a 
momentum cutoff at $|k|=\s$ there is an intrinsic short-distance cutoff, on the order
$l=\pi/\s$, on the spatial variation of $f^{low}(x)$.  Therefore, the dimensional reduction form for $F_{12}^{low,a}$
only implies that the correlation length, extracted, e.g., from the correlator 
\beq
        \langle \mbox{Tr}[(F_{12}^{low}(x))^{2}] \mbox{Tr}[(F_{12}^{low}(y))^{2}] \rangle 
             - \langle \mbox{Tr}[(F_{12}^{low}(x))^{2}] \rangle^{2}
\eeq
is less than $l$.   This upper bound on the correlation length is reduced as $\s$ is increased,
until, at  $\s \approx \pi/l_{K}$, the Fourier modes $S(k)$ at $|k|<\s$ cease to be constant,  and $f^{low}(x)$
does not fluctuate independently in regions separated by distances $l<l_{K}$.  Thus the range
$l_{K}$ of the vacuum kernel is also an estimate of the correlation length, which is the inverse
of the mass gap. Denoting $m_{K}=l_{K}^{-1}$, we would then have $m_{K}\approx m_{0^{+}}$, 
where $m_{0^{+}}$ is the mass of the $0^{+}$ glueball. 
 
   At first sight, however, there would seem to 
be no reason that $1/\sqrt{-\D^{2}}$ should have a finite range.\footnote{For this reason it was suggested by Samuel, 
in ref.\ \cite{Samuel} that one could introduce a variational parameter $m^{2}$ into the kernel, i.e.\
$1/\sqrt{-\D^{2}+m^{2}}$, as a way of introducing a finite range.  The value of $m^{2}$ 
would be determined, in principle, by minimizing the energy expectation value of the trial vacuum state.}
But in fact, for the covariant Laplacian in two Euclidean dimensions, there is good reason to believe
that the corresponding Green's function is finite range for almost
any stochastic background $A_{\m}$ field, no matter how weak.   The relevant phenomenon is known as
\emph{weak localization} \cite{weaklocal}.   Consider a non-relativistic particle moving in a stochastic potential
in one or two space dimensions.  It is known that the eigenstates of the corresponding Hamiltonian ${\cal H}$ 
are all localized; i.e.\ they fall off exponentially outside some finite region, no matter how weak the stochastic potential
may be.   In that case, the inverse operator ${\cal H}^{-1}$ has a finite range.   Weak localization is
expected to occur also in cases where the stochasticity lies in the kinetic, rather than the potential term, and 
this seems to be relevant to the case at hand, i.e.\ the two dimensional covariant Laplacian.  If all of the eigenmodes
of $-\D^{2}$ are localized, then both $1/(-\D^{2})$ and $1/\sqrt{-\D^{2}}$ are necessarily finite range.
In fact, the range is finite even if there is only an interval of localized eigenmodes at the low end of the 
spectrum, with the bulk of the states non-localized \cite{Us1,Stone}. 

       The range of the vacuum kernel $1/\sqrt{-\D^{2}}$ is fixed by self-consistency.  In two dimensions there
is no Bianchi identity, so a correlation between field strengths at different points in space, at equal times,
can only arise from the non-local character of the vacuum wavefunctional.  As argued above, field-strength correlations are 
negligible on a scale at which the vacuum wavefunctional has the form \rf{eff}, and the wavefunctional \rf{state} takes
this form on a scale which is roughly equal to the range of the vacuum kernel.  This means that the correlation length 
of field strengths in a typical vacuum fluctuation (i.e.\ the inverse of the $0^{+}$ glueball mass), should approximate 
the range of the kernel $1/\sqrt{-\D^{2}}$,  but the range of the kernel, in turn, depends on the vacuum fluctuation.

     The first question is whether, in a vacuum fluctuation typical of massless free field theory, having infinite range
correlations among field strengths, the kernel can also have an infinite range.  If the kernel has a finite range 
in such a background, then massless free field fluctuations are inconsistent; they do not arise from the behavior of
the kernel that they imply.   We check this behavior numerically in the following way:  It is required to generate 
stochastically a gauge field background that would arise in a free theory, from a probability distribution
$|\Psi_{0}^{free}[A]|^{2}$, where, in momentum space,
\beq
           \Psi^{free}_{0}[A] = \exp\left[-\oh \int d^{2}k  ~k A^{T}_{i}(k) A^{T}_{i}(-k) \right]
\eeq    
The superscript $T$ means ``transverse".  In this initial investigation the transversality
restriction is ignored, and the continuum is replaced by a lattice.   Let $-\nabla^{2}$ be the (ordinary) lattice
Laplacian, with eigenstates $\{\varphi_{n}(x)\}$ and eigenvalues $\{\l_{n}\}$, and let $\{s_{n,i}\}$ be a set of
random numbers taken from a normal distribution with unit variance.  
Then a typical free-field vacuum fluctuation, deriving from the probability distribution 
$|\Psi^{free}_{0}[A]|^{2}$ (with the transversality condition dropped) is
\beq
          A_{i}(x)  = \sum_{n=2}^{L^{2}} {s_{n,i}\over \l_{n}^{1/4}} \varphi_{n}(x)
\label{sum}
\eeq
This construction is repeated three times, for each of the three color indices, to form the field 
$A_{i}^{a}(x)$.  

    Although no coupling constant is used to generate the free $A$-fields, there is still a Yang-Mills coupling
constant present in the definition of the covariant derivative.  In lattice regularization, this coupling constant 
relates the $A$-fields to the SU(2) link variables 
\beq
          U_{i}(x) = \exp\left[ ig\sum_{c=1}^{3} A_{i}^{c} \oh \sigma^{c} \right]
 \label{link}
 \eeq
 From the link variables in the fundamental 
 representation, we then obtain link variables ${}^{A}U$ in the adjoint representation 
 \beq
          {}^{A}U^{ab}_{i}(x) = \oh \mbox{Tr}[\s_{a} U_{i}(x)\s_{b}U^{\dg}_{i}(x)]
 \eeq 
 The covariant lattice Laplacian, in the adjoint representation, is then  
 \beq
          \Bigl(\D^{2}\Bigr)^{ab}_{xy} = \sum_{k=1}^{2} \Bigl[ {}^{A}U^{ab}_k(x) \d_{y,x+\hat{k}}
         + {}^{A}U^{\dg ab}_k(x-\hat{k}) \d_{y,x-\hat{k}}  - 2 \d^{ab} \d_{xy} \Bigl]
 \eeq
 Combining color and space indices, the covariant lattice Laplacian
 can be regarded as a sparse $3L^{2} \times 3L^{2}$ matrix.   Standard numerical routines 
 are used to take the inverse square root of this matrix, to arrive at 
 \beq
           G^{ab}(x,y) = \left({1\over \sqrt{-\D^{2}}}\right)^{ab}_{xy}
 \eeq
To calculate the range, we form the gauge-invariant combination
\beq
           \widetilde{G}(x,y) = \sqrt{\sum_{a,b} G^{ab}(x,y) G^{ba}(y,x)}
\eeq
and average $\tilde{G}(x,y)$ over all points $x,y$ on the lattice with fixed $|x-y|=R$.  Denote the result $G(R)$.  
A linear fit to $\log[G(R)]$ vs.\ $R$ determines the inverse $m_{K}$ of the range $l_{K}$ of the kernel.

\begin{figure}[t!]
\includegraphics[width=8.8truecm]{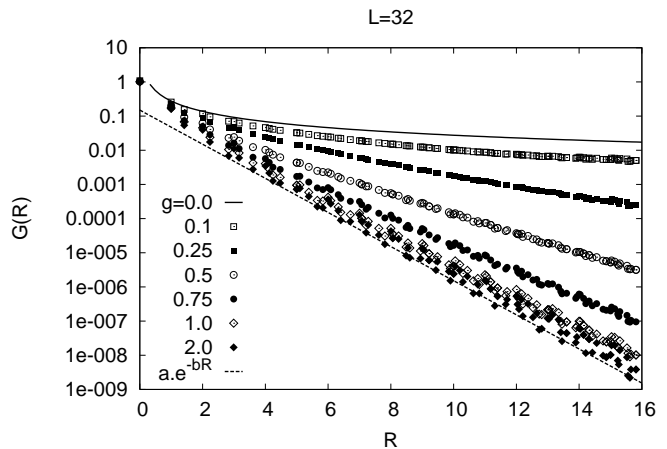}
\caption{The modulus of the vacuum kernel $1/\sqrt{-\D^{2}}$ in a ``free field" background, but
with various non-zero values of the Yang-Mills coupling $g$ in the covariant Laplacian.  
The upper solid line is the kernel $G(R)=\sqrt{3}/(2\pi R)$ corresponding to $A^a_i(x)=0$ (or, equivalently, $g=0$).
Data at high $g$ tends to fall near the lower dashed line.}
\label{task3}
\end{figure}
  
     The results, for couplings from $g=0.1$ to $g=2.0$ on a $32^{2}$ lattice are shown in Fig.\  \ref{task3},
which plots $G(R)$ vs.\ $R$ on a semilog scale.  
The solid line is the result for $g=0$ in the continuum, which is simply $G(R) = \sqrt{3}/(2\pi R)$.
Even at $g=0.1$, where the average link variable $\langle \oh \mbox{Tr}[U] \rangle = 0.997$ 
is very close to unity, there is a noticeable deviation from $1/R$ behavior, and 
at all the larger $g$ values the $32^{2}$ lattice is big enough to clearly see an exponential
falloff in $G(R)$.\footnote{We note that the data at $g>1.0$ also approaches a limiting line, approximately at
the position of the lower line in Fig.\ \ref{task3}.  This simply means
that, in the way we have generated the link variables, taking $g\ra \infty$ does not result in a random link
distribution equal to the Haar measure.}

    This data indicates that $1/\sqrt{-\D^{2}}$ is finite range, even in a massless free-field
background. This implies that free fields, and probably any sort of non-confining vacuum fluctuations
in $2+1$ dimensions, are inconsistent.  The kernel is finite range in general, dimensional reduction
is obtained, and the simple 
vacuum wavefunctional \rf{state} is confining.

    Given that the approximate ground state wavefunctional $\Psi_{0}$ of eq.\ \rf{state} is confining, 
the next question is how closely this state approximates the true vacuum state of SU(2) Yang-Mills theory. 
We have already argued that in the state \rf{state}, $m_{K}\approx m_{0^{+}}$.   If $\Psi_{0}$ is a good 
approximation to the true Yang-Mills vacuum
state $\Psi_0^{true}$, then in the true vacuum state $m_{K}$ and $m_{0^+}$ should still be approximately equal.
To test this approximation, we calculate the
range of the kernel $1/\sqrt{-\D^{2}}$ in a 2D time-slice of a thermalized lattice, generated
by Monte Carlo simulation of lattice Yang-Mills theory in D=3 Euclidean dimensions.   This procedure
makes use of the fact that
if $O[A]$ is an observable defined at fixed time $t$, then
\bea
           \langle O \rangle &=& {1\over Z} \int DA ~ O[A] e^{-S}
\non \\
                  &=& \langle \Psi^{true}_{0}|O|\Psi^{true}_{0} \rangle
\eea
Comparison of $m_{K}$ with the known lattice results for the mass of the $0^{+}$ glueball
is one measure of how well the zeroth-order state $\Psi_{0}[A]$ approximates the true ground state, since if
$\Psi^{true}_{0}=\Psi_{0}$, then $m_{K}\approx m_{0^{+}}$.   The procedure for calculating $m_{K}$ in the true vacuum
is as follows:  Link variables are generated by standard lattice Monte Carlo of $D=3$ dimensional lattice gauge theory
with a Wilson action, and the covariant Laplacian is evaluated in a time-slice.  From this one calculates 
$1/\sqrt{-\D^{2}}$ and $G(R)$, as described above.  Then $\log G(R)$ is fit to a straight line in the
range $R\in [{L\over 4},{L\over 2}]$; the kernel mass $m_{K}$ is just $-1$ times the slope.  The procedure is repeated for
a number $(>10)$ of independent thermalized lattice configurations, and the results are averaged.
      
\begin{figure}[t!]
\includegraphics[width=8.8truecm]{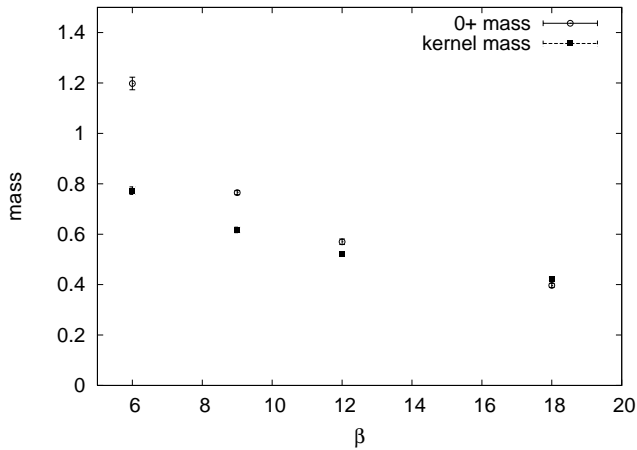}
\caption{A comparison, at various lattice couplings $\b$ in $D=2+1$ dimensions, 
of the inverse range $m_{K}$ of the vacuum kernel, and the mass of $0^{+}$ glueball reported 
in ref.\ \cite{Teper}.  Errorbars are comparable to symbol size.}
\label{task5}
\end{figure}
    
        The values of $m_{K}$ obtained in this way at a number of different lattice couplings $\b$ are shown in 
Fig.\ \ref{task5}, where we have also displayed the $0^{+}$ glueball masses that have been computed 
via lattice Monte Carlo  by Meyer and Teper in ref.\ \cite{Teper}.  For the evaluations of $m_{K}$
we have used lattice extensions $L=16,24,32,40$ 
at $\b=6,9,12,18$ respectively.  With the exception of $\b=18$, where Meyer and Teper used a slightly larger
lattice ($L=50$), these are the same lattice volumes used in ref.\ \cite{Teper}.   In the range of $\b$ studied,
$m_{K}$ and $m_{0^+}$ appear to converge as $\b$ increases.   This is the behavior one would expect, if the true
vacuum and the zeroth-order approximation \rf{state} to the ground state approach one another in the weak 
coupling, $\b \ra \infty$ limit.

      Despite this numerical success, it cannot be true that the zeroth-order state and the true ground state agree
in all essential ways.   Since the zeroth-order state has the dimensional reduction form at large scales,  the dependence 
of the string tension on the group representation of the quark-antiquark sources is the same as in $D=2$ Euclidean
dimensions, i.e.\ Casimir scaling.  While this dependence is correct at intermediate distance scales, and also
asymptotically in the $N\ra \infty$ limit, it just cannot be true asymptotically at finite $N$, where
the asymptotic string tension depends only on the $N$-ality of the quark sources.  The simple zeroth-order
state appears to be missing the center vortex (or domain) structure, which has to dominate the vacuum state
at sufficiently large scales \cite{review,g2}.   So dimensional reduction is not the whole story at finite $N$,
and higher-order, $1/N^{2}$ suppressed terms in the vacuum wavefunctional must come into play at 
sufficiently large distance scales.  At present we have no systematic method for determining these higher-order terms.

     It should be noted that there have been a number of efforts to determine the Yang-Mills vacuum wavefunctional
beyond weak-coupling perturbation theory.  These are mainly variational in character; see in particular
\cite{Adam,Alex}.   A non-variational approach is followed by Leigh et al.\ \cite{Rob} in 
their recent work on the Yang-Mills ground state  in $2+1$ dimensions.  There are some strong similarities 
between that work and the present paper.  Leigh et al.\ arrive, by various manipulations, at a ground state
which is the exponential of a bilinear expression, involving a finite range kernel connecting 
gauge-invariant variables related to the field strength.  Due to  
the finite range of the kernel, the vacuum state derived in ref.\ \cite{Rob} should have the dimensional 
reduction form \rf{eff} when viewed at large scales.  This, of course, is very reminiscent of $\Psi_{0}$ in eq.\ \rf{state}.  
However, it is hard to compare these vacuum states directly.  For one thing, the two states are expressed in terms of 
different variables; ref.\ \cite{Rob} makes use of the Karabali-Nair variables in 2+1 dimensions.  
For another, the kernel of ref.\ \cite{Rob} is gauge-invariant, 
rather than gauge-covariant,  and has a finite range even for a vanishing background field.  That is not at all like 
the behavior of the kernel $1/\sqrt{-\D^{2}}$.  Still, there are enough superficial similarities between these two states 
to make one suspect that they are related.

    We would like to conclude with some remarks about 3+1 dimensions.   The evidence presented above
suggests that confinement in 2+1 dimensions is simply a consequence of the physical state constraint 
(i.e.\ gauge invariance), plus the free-field limit of the vacuum wavefunctional.  The key point
is that gauge invariance introduces a gauge covariant kernel $1/\sqrt{-\D^{2}}$, and this kernel, unlike
$1/\sqrt{-\nabla^{2}}$, has a finite range in almost any stochastic background, possibly due to the weak 
localization property in two Euclidean dimensions.    The interesting question is whether
the gauge-covariant kernel still has a finite range, in a free-field background at fixed time, in the 3+1 dimensional 
theory.  It is known that the spectrum of $-\D^2$ has an interval of localized states in $D=4$ Euclidean dimensions, 
in a confining background \cite{Us1,Us2}.  But it is not known whether this operator has localized states in a free-field 
background in $D=3$ dimensions, or else whether the covariant kernel at $D=3$ is finite range for some other reason.  
If the kernel is, in fact, finite range, then on large scales $|\Psi_{0}|^{2}$ effectively 
reduces to the probability distribution of a three Euclidean dimensional Yang-Mills theory, and the conjectured 
two-step dimensional reduction of eq.\ \rf{dimred}, resulting in an area law falloff for large planar Wilson loops, 
would hold.  The matrix operations would have to be carried out on $3L^{3} \times 3L^{3}$ 
matrices, rather than the $3L^{2} \times 3L^{2}$ size that appears in $2+1$ dimensions.  This additional computation cost, 
while restrictive, is probably not prohibitive. 
   
\acknowledgments{
     We thank Dmitri Diakonov, Robert Leigh, Adam Szczepaniak, 
and Valentin Zakharov for helpful discussions. This research is supported in part by 
the U.S.\ Department of Energy under 
Grant No.\ DE-FG03-92ER40711 (J.G.), the Slovak Science and Technology 
Assistance Agency under Contract No.\ APVT--51--005704 (\v{S}.O.), 
and the Grant Agency for Science, Project VEGA No.\ 2/6068/2006 (\v{S}.O.).
}

\end{document}